\documentclass{webofc}
\usepackage[varg]{txfonts}   % Web of Conferences font

\usepackage[normalem]{ulem}
\usepackage{amsmath, amsfonts, graphicx,color,slashed,bm,fixmath}

\newcommand{\be}{\begin{equation}}
\newcommand{\ee}{\end{equation}}
\newcommand{\bea}{\begin{eqnarray}}
\newcommand{\eea}{\end{eqnarray}}

\begin{document}

\title{Inhomogeneous pion condensed phase hosting topologically stable baryons} 
%\title{ Topological solitons with baryonic charge in an inhomogeneous pion gas}

\author{\firstname{Massimo} \lastname{Mannarelli}\inst{1} \and
        \firstname{Fabrizio} \lastname{Canfora}\inst{2} \and
        \firstname{Stefano} \lastname{Carignano}\inst{3} \and  \firstname{Marcela} \lastname{Lagos}\inst{4}\and
         \firstname{Aldo} \lastname{Vera}\inst{4} \fnsep\thanks{\email{massimo@lngs.infn.it}\\ \email{canfora@cecs.cl}\\ \email{carignano@ice.cat}\\ \email{marcela.lagos@uach.cl} \\ \email{aldo.vera@uach.cl}}
        % etc.
}

\institute{INFN, Laboratori Nazionali del Gran Sasso, Via G. Acitelli,
22, I-67100 Assergi (AQ), Italy 
\and
Centro de Estudios Cient\'{\i}ficos (CECS), Casilla 1469,
Valdivia, Chile
\and
Departament de F\'isica Qu\`antica i Astrof\'isica and Institut
de Ci\`encies del Cosmos, Universitat de Barcelona, Mart\'i i Franqu\`es 1,
08028 Barcelona, Catalonia, Spain
           \and
           Instituto de Ciencias F\'isicas y Matem\'aticas, Universidad
Austral de Chile,  Casilla 567, Valdivia, Chile}

\abstract{ We discuss the inhomogeneous pion condensed phase within the framework of chiral perturbation theory. We show how the  general expression of the condensate can be obtained solving three coupled   differential equations,  expressing how the  pion fields are modulated in space.   Upon using some simplifying assumptions, we determine an analytic solution in (3+1)-dimensions. The obtained  inhomogeneous condensate is characterized by a non-vanishing topological charge, which can be identified with the  baryonic number. In this way, we  obtain an inhomogeneous system of  pions hosting  an arbitrary number of baryons at fixed positions in space.}
\maketitle
\section{Introduction}
\label{sec:intro}
One of the main goals of the high-energy physics community is to determine the properties of hadronic matter at extremes of temperature and baryonic densities. The fundamental motivation behind these studies  is that the high-energy environment is the ideal one for probing  the constituents of hadronic matter. At the microscopic level, the strong interaction  is  described by Quantum Chromodynamics (QCD), which is an extremely challenging  theory in the infrared, because at low energies the basic constituents, quarks and gluons, are confined.  On the other hand, with increasing energy scale we expect that quarks and gluons are liberated~\cite{Cabibbo:1975ig},  and that perturbative methods can be eventually used at asymptotic energy scales. Unfortunately, in terrestrial relativistic heavy-ion colliders, as well as in compact stellar objects, the energy scale is not sufficiently large to ensure that perturbative methods are under control. This means that different phenomenological methods should be used to infer the properties of hadronic matter. 

\begin{figure}
\centering
\sidecaption
\includegraphics[width=9.5cm]{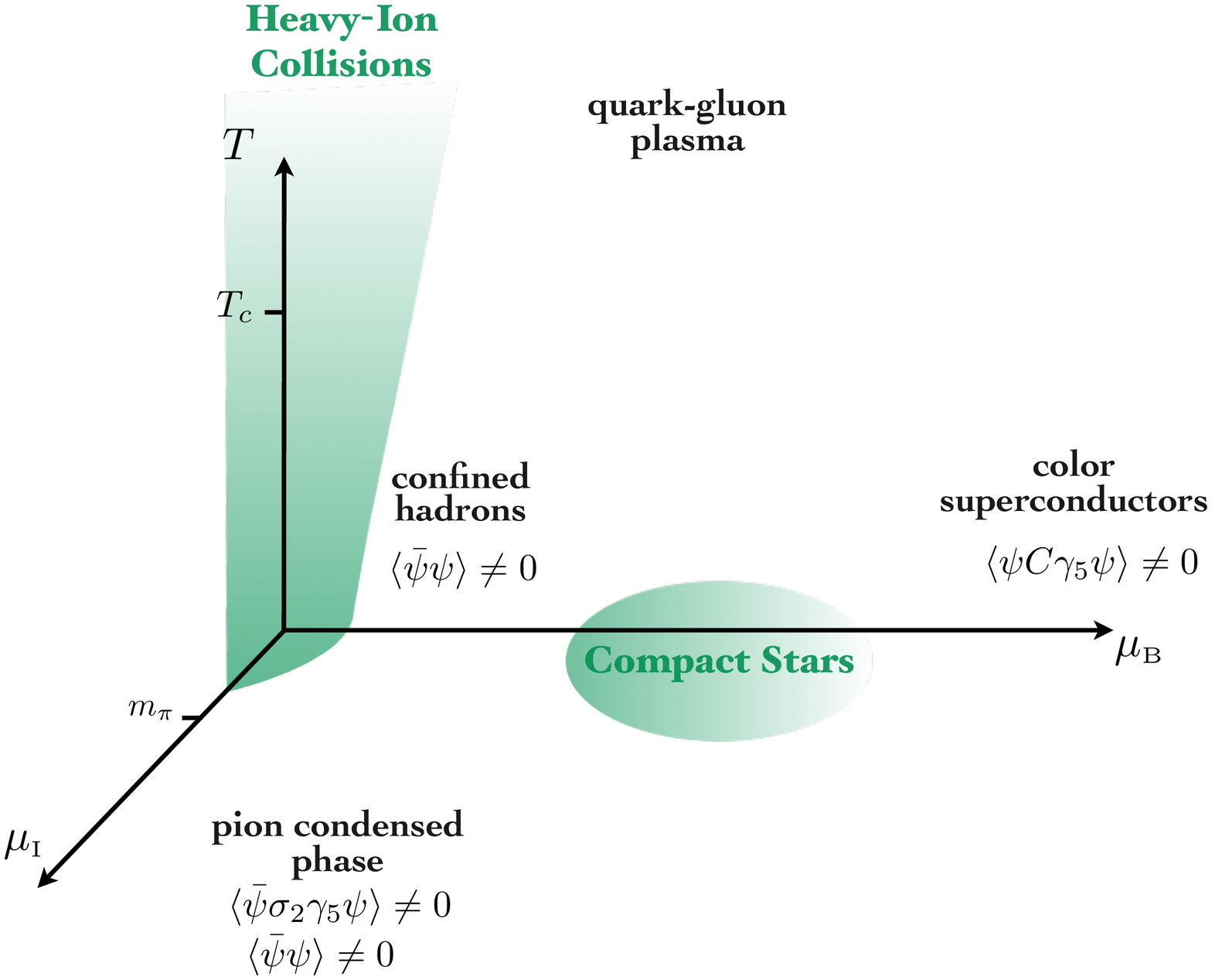}
\caption{Cartoon of the QCD phase diagram. A grand-canonical description of the phases of matter obtained by changing the temperature, the baryonic chemical potential and the isospin chemical potential. The relevant di-quark condensates are shown. }
\label{fig-1}       % Give a unique label
\end{figure}

The possible  phases of hadronic matter  can be depicted in a grand-canonical phase diagram, which we pictorially report in Fig.~\ref{fig-1}. The phases of matter change as a function of temperature, $T$, baryonic chemical potential, $\mu_B$, and isospin chemical potential, $\mu_I$, and can be characterized by different values of the di-quark condensates.
  
In Fig.~\ref{fig-1} are also shown the regions which roughly correspond to those accessible by relativistic heavy-ion collisions and compact stars.
Although weak decays typically determine the isospin chemical potential at any temperature and baryonic chemical potential, it is useful to assume that $\mu_I$ is a free parameter. In this way one can explore the properties of matter at short time scales, that is shorter than the  typical time-scale of weak interactions. Moreover, when the properties of hadronic matter  are studied in lattice QCD simulation, it is typically assumed that the weak interaction is turned off. 

Although lattice QCD simulations at non-vanishing $\mu_B$ are hampered by the so-called sign problem, they are feasible at non-vanishing isospin chemical potential~\cite{Alford:1998sd}. This offers the opportunity to study deconfinement as a function of $\mu_I$ and to compare the results with the outcome of effective field theories, in particular with  the results of chiral perturbation theory  ($\chi$PT), see ~\cite
{Weinberg:1978kz, Gasser:1983yg, Georgi:1985kw, Leutwyler:1993iq,
Ecker:1994gg, Leutwyler:1996er, Pich:1998xt, Scherer:2002tk, Scherer:2005ri}, which is a low-energy realization of hadronic matter having  the same  global symmetries of QCD.  As an effective field theory, $\chi$PT perfectly describes the properties of pseudoscalar mesons, and it is therefore particularly useful for analyzing matter at vanishing $\mu_B$ and non-zero $\mu_I$. In principle, $\chi$PT can be amended to describe baryons as well, see for instance  \cite{Pich:1998xt, Scherer:2002tk, Scherer:2005ri}, introducing the baryonic sources by hand.  In~\cite{Canfora:2020uwf} we have  presented  a  different approach, developed with the aim of obtaining  baryons  as skyrmions embedded in an inhomogeneous gas of pions at non-vanishing $\mu_I$. In this way, baryons emerge as solitonic solutions of the classical field equations. Since the obtained set of differential equations  is quite difficult to solve, we have used a number of simplifying assumptions that make the problem almost analytically tractable. 

\section{Moving along the $\mu_I$--axis}
\label{sec:chipt}
 In $\chi$PT the three pions can be collected in a
unimodular field 
\begin{equation}  \label{eq:Sigma}
 \Sigma  ={\bm 1}_{2} \cos \alpha + i\,
 \mathbf{n} \cdot \mathbold{\sigma} \sin \alpha\,,
\end{equation}
where the components of $\mathbold{\sigma}$ are the three Pauli matrices, and  the three  degrees of freedom are $\alpha$  and the two independent components of the  unimodular field $\bm n$.  
This somehow unusual representation of the pion fields  is particularly  convenient for the analysis of different ground states. As we shall see, different vacua  are indeed characterized by different values of the tilting angle $\alpha$. 

In $\chi$PT the interaction terms are grouped using an expansion in
exchanged momenta. The leading order  $\chi$PT Lagrangian including the electromagnetic
field can be written as 
\begin{align}
\mathcal{L}& =\frac{f_{\pi }^{2}}{4} \mathrm{Tr}\left[- \left( \Sigma ^{\mu
}\Sigma _{\mu }\right) + m_{\pi }^{2}\left( \Sigma +\Sigma ^\dagger\right) %
\right] -\frac{1}{4}F_{\mu \nu }F^{\mu \nu } \,,  \label{eq:lagrangian}
\end{align}
where the field strength is $F_{\mu \nu }=\partial _{\mu }A_{\nu }-\partial _{\nu
}A_{\mu }$, with  $A_\mu $ the electromagnetic field, and 
\begin{align}  \label{eq:sigmamu}
\Sigma _{\mu } =\Sigma^{-1}D_{\mu }\Sigma\,,
\end{align}
with 
\begin{equation}\label{eq:covariant}
D_{\mu }\Sigma=\partial_{\mu }\Sigma+i \widetilde{A}_{\mu }\left[
\sigma_{3},\Sigma\right] \ ,
\end{equation}%
the covariant derivative. Here 
\begin{equation}
\widetilde{A}^{\mu }=\left( \frac{\mu _{I}}{2}+A^{0},\bf{A}\right) \ ,
\label{eq:deftilda}
\end{equation}%
is a convenient definition of  the shifted electromagnetic field that includes the effect of the isospin chemical potential. 
The low-energy phenomenological constants appearing in  Eq.~\eqref{eq:lagrangian} are  the pion decay constant,  $f_{\pi } \simeq 93 $ MeV, and the pion mass, $m_{\pi } \simeq 135$ MeV; we assume 
degenerate pion  masses.

The homogeneous energetically favored phase can be determined assuming that $\alpha$ and $\bm n$ are variational parameters: the favored state corresponds to the values  of these parameters that maximize the Lagrangian in \eqref{eq:lagrangian}, see \cite{Mannarelli:2019hgn} for a review. It results that for $|\mu_I| < m_\pi $ the standard vacuum with $\alpha=0$ is energetically favored. In this case, the value of the  vector $\bm n$ is not specified, and the only non-vanishing condensate is the so-called chiral condensate
\be
\sigma  \propto \langle \bar \psi \psi \rangle\,,
\ee
which in terms of the quark fields $\psi$ corresponds to the locking of  the chiral symmetry SU(2)$_L\times$ SU(2)$_R $  to the vector SU(2)--isospin symmetry. Upon increasing $\mu_I$ the isospin symmetry is explicitly broken and the masses of the charged pions split, similarly to what happens with the Stark effect. However, the system has still a $U(1)$ global symmetry. When $|\mu_I| > m_\pi $ this symmetry 
is spontaneously broken, one the pion field becomes massless and the system becomes superfluid. In this case the vacuum is \emph{tilted} in isospin space by an angle 
\be\label{eq:alpha_pi}
    \cos \alpha=\left(\frac{m_\pi}{\mu_I}\right)^2\,,
    \ee  
corresponding to the fact that in addition to the chiral condensate, the pion condensate
\be
 \pi  \propto \langle \bar \psi \sigma \gamma_5 \psi \rangle\,,\label{eq:pion_cond}\\
\ee
assumes a non-vanishing  value. The energetically favored configuration is such that the third component of the field $\bm n$ vanishes. The remaining two components of $\bm n$ are not determined, and the only remaining independent degree of freedom corresponds to the  Nambu-Goldstone field. When considering the effect of the electromagnetic interaction, the  system becomes a superconductor and the electromagnetic interaction is screened, see~ \cite{Mammarella:2015pxa,Mannarelli:2019hgn} for more details.

\section{Inhomogeneous phase}
\label{sec:sec-2}
\begin{figure}[h!]
\centering
\sidecaption
\includegraphics[width=9.5cm]{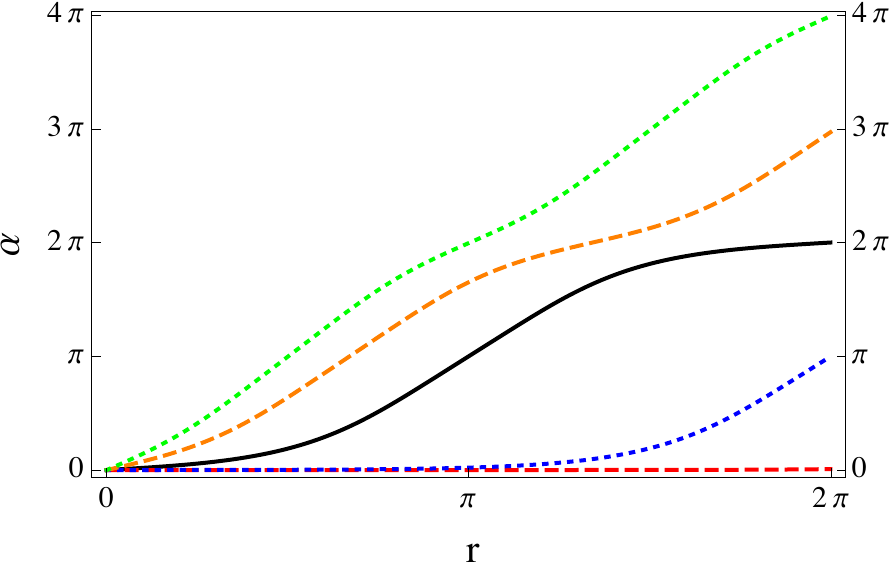}
\caption{Profile of the radial mode, $\alpha(r)$, obtained
by numerical integration of Eq.~\eqref{eq:alpha_sol} for $q=1$ and $\ell=1/m_\pi$ and assuming the boundary conditions  in Eq.~\eqref{eq:BCalpha}. The reported  results are obtained with
five different values of the winding number $n$. }
\label{fig:alpha}
\end{figure}

\begin{figure*}[thb!]
\includegraphics[width=6cm]{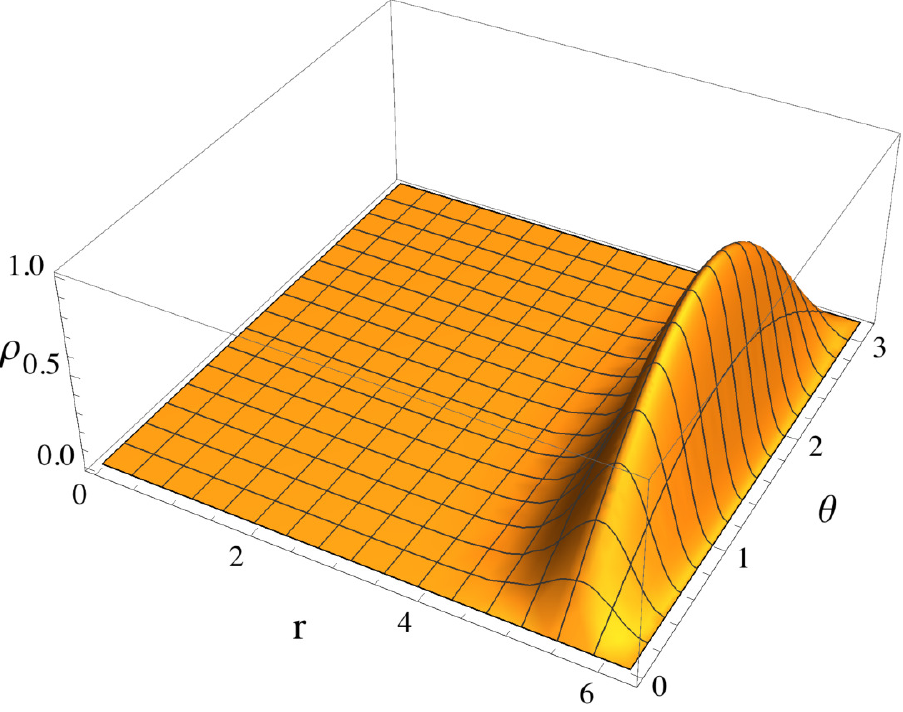} \hspace{1.cm} %
\includegraphics[width=6cm]{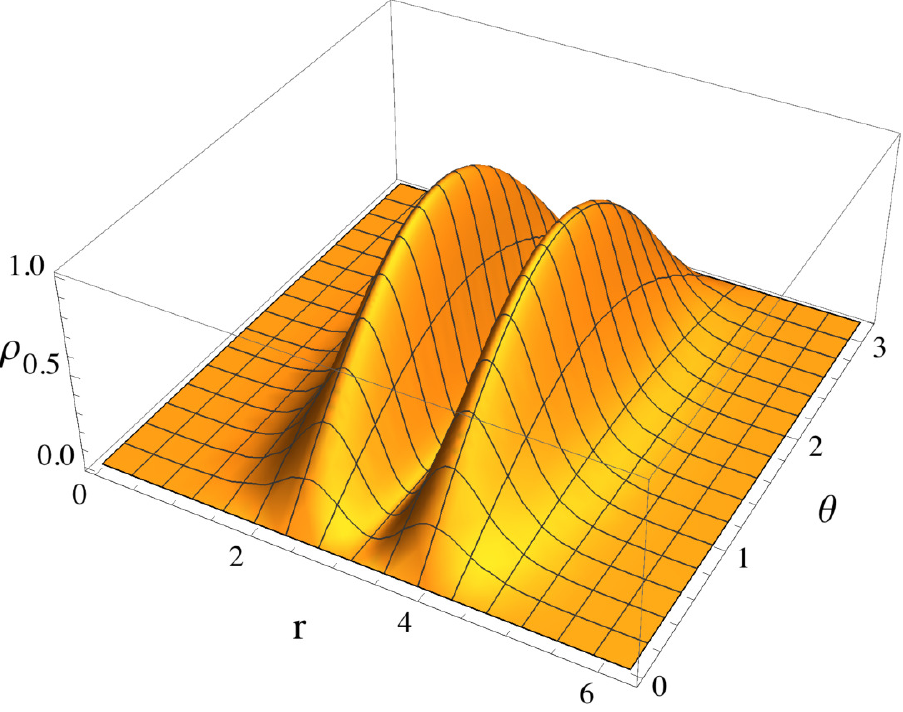}
\caption{Topological density distribution, see Eq.~\eqref{eq:rhom2}, obtained with $q=1$ and  $n=1$, left panel, and $n=2$, right panel. The topological density has been normalized to the
value at the maximum. }
\label{fig:solitons}
\end{figure*}

The picture emerging so far is that of a homogeneous system that reacts to an isospin chemical potential exceeding the 
pion mass by the appearance of a pion condensate. This condensate  coexists  with  the chiral condensate.
In this situation, it is possible that the energetically favored configuration is such that the space distribution of the two condensates is not uniform. To study this possibility, we have to promote the variational parameters $\alpha$ and $\bm n$ to classical fields. For vanishing electromagnetic fields, the  effective Lagrangian in Eq.~%
\eqref{eq:lagrangian} can be rewritten as 
\begin{align}  \label{eq:Lgen}
\mathcal{L}_\text{m} = & \frac{f_\pi^2}2 \left[ \partial_\mu \alpha
\partial^\mu \alpha +\sin^2{\alpha} \partial_\mu \Theta \partial^\mu \Theta
+ 2 m_\pi^2 \cos\alpha \right.  \notag \\
&\left. +\sin^2{\alpha} \sin^2{\Theta}(\partial_\mu \Phi -\mu_I \delta_{\mu
0})(\partial^\mu \Phi-\mu_I \delta^{\mu 0}) \right]\,,
\end{align}
where we have parameterized the components of $\bm n$ as follows
\begin{align}
n_{1}& =\sin \Theta \cos \Phi \ ,\ \ n_{2}=\sin \Theta \sin \Phi \ ,\ \
n_{3}=\cos \Theta \,.  \label{eq:unitvector}
\end{align}
The three classical fields $\alpha, \Theta$ and $\Phi$ can be obtained as solutions of the Euler-Lagrange equations
\begin{align}
\partial_\mu\partial^\mu\Phi =& - (\partial_\mu\Phi - \mu_I \delta_{\mu_0})
\partial^\mu (\log(\sin^2\alpha \sin^2\Theta) )\,,  \label{eq:Phi} \\
\partial_\mu\partial^\mu\Theta =& -2 \cot
\alpha\,\partial^\mu\Theta\partial_\mu\alpha+ \frac{\sin2\Theta}{2} K \,,
\label{eq:Theta} \\
\partial_\mu \partial^\mu \alpha = & - m_\pi^2 \sin\alpha + \frac{\sin(2
\alpha)}{2}( \partial_\mu \Theta \partial^\mu \Theta + K \sin^2\Theta )\,,
\label{eq:alpha}
\end{align}
 which is a set  of three coupled differential
equations, with  \be\label{eq:K} K= (\partial_\mu\Phi - \mu_I \delta_{\mu_0})(\partial^\mu\Phi - \mu_I
\delta^{\mu_0})\,.
\ee
Since we are interested in  topologically protected  phases, it is mandatory to take into account finite volume
effects. For this reason, we define the line element as 
\begin{equation}
ds^{2}=dt^{2}- \ell^2\left( dr^{2}+d\theta ^{2}+d\phi ^{2}\right) \ ,
\label{Minkowski}
\end{equation}%
where the  coordinates $r$, $\theta $ and $\phi $
vary in the ranges 
\begin{equation}
0\leq r\leq 2\pi \ ,\quad 0\leq \theta \leq \pi \ ,\quad 0\leq \phi \leq
2\pi \,,  \label{eq:ranges}
\end{equation}%
while
\begin{equation}  \label{eq:ell}
\ell=\frac{b}{m_\pi} \,,
\end{equation}
with $b$ a real number,  is  the typical dimension of the system. The actual solution is expected to satisfy some appropriate Dirichlet boundary conditions. In particular,
we demand that 
\begin{equation}  \label{eq:BC1}
n(t,r,0,\phi) = - n(t,r,\pi,\phi)\,, \quad n(t,r,\theta,0) = n(t,r,\theta, 2
\pi) \ ,
\end{equation}
and that 
\begin{equation}  \label{eq:BC2}
\Sigma(t,0,\theta,\phi) = \pm \Sigma(t,2\pi,\theta,\phi)\,,
\end{equation}
where we have taken the freedom to have a  flip of the $\Sigma$ field at the boundary, see \cite{Canfora:2018rdz, Canfora:2019kzd, Canfora:2020kyj, Canfora:2020uwf} for more details. Since solving the general Cauchy problem is hard, we take the  simplifying assumption that
\begin{equation}
\Phi =\mu_I \ell \phi + \Phi_0 \ ,\qquad \Theta =q \theta + \Theta_0 \,,
\label{eq:general}
\end{equation}
with $q$ odd and \be\label{eq:p} \mu_I \ell =p \,, \ee with $p$ an integer, to properly satisfy the above  boundary conditions,
see~\cite{Canfora:2018rdz, Canfora:2019kzd, Canfora:2020kyj, Canfora:2020uwf}. The particular choice of the $\Phi$ field is such that $K$ vanishes, see Eq.~\eqref{eq:K}. In this way, the only nontrivial differential equation turns to be
\begin{align}
\frac{\partial^2 \alpha}{\partial r^2} = & m_\pi^2 \ell^2 \sin\alpha + \frac{%
q^2}{2} \sin(2 \alpha)\,,  \label{eq:alpha_sol}
\end{align}
where we have assumed that $\alpha$ is stationary. The appropriate  boundary conditions in the $r$--direction are then
\begin{equation}  \label{eq:BCalpha}
\alpha(0) =0 \qquad \text{and} \qquad \alpha(2 \pi) = n \pi\,,
\end{equation}
with $n$ the integer winding number. Odd values of $n$ correspond to the minus sign in  Eq.~\eqref{eq:BC2}, while even values of $n$ correspond to a positive sign, that is anti-periodic and periodic boundary condition in the $r$--direction, respectively.  Some solutions of this differential equation are shown in Fig.~\ref{fig:alpha}, for five different winding numbers.
The obtained spatial modulations are characterized by a non-vanishing topological charge. This charge  can indeed be written as 
\begin{equation}
B_\text{m}=\frac{\ell^3}{24\pi ^{2}}\int_V drd\theta d\phi \ \rho _{\text{m}}\
\,,  \label{eq:topo1}
\end{equation}%
where $V$ is the considered volume and
\begin{equation}  \label{eq:rhom}
\rho _{\text{m}}=\epsilon ^{ijk}\text{Tr}\left\{ \left( \Sigma ^{-1}\partial
_{i}\Sigma \right) \left( \Sigma ^{-1}\partial _{j}\Sigma \right) \left(
\Sigma ^{-1}\partial _{k}\Sigma \right) \right\} \ ,
\end{equation}%
is the topological density of the matter fields. The expression
in Eq.~\eqref{eq:rhom} depends on the space derivatives of the fields, and it vanishes if the scalar field $%
\Sigma $ depends on only one or two space coordinates. Upon substituting Eq.~\eqref{eq:Sigma} in Eq.~%
\eqref{eq:rhom}, the topological density can be rewritten as 
\begin{align}
\rho_{\text{m}} =-12 \frac{pq}{\ell^3}\sin (q\theta )\sin ^{2}(\alpha ) \frac{\partial\alpha}{\partial r}\,,
\label{eq:rhom2}
\end{align}
which can be readily integrated using the fact that
\begin{equation}
\int_0^{2\pi} dr \sin ^{2}(\alpha) \frac{\partial\alpha}{\partial r} = \int_0^{n \pi} d\alpha \sin^2 \alpha = n \frac{\pi}2\,.
\label{eq:alpha_int}
\end{equation}
According to the boundary condition in Eq.~\eqref{eq:BCalpha},
the topological charge takes the values
\begin{equation}
B_{\text{m}}=%
\begin{cases}
n p & \text{if}\quad q\text{ odd} \ , \\[1.5ex] 
0 & \text{if}\quad q\text{ even} \ ,%
\end{cases}
\label{eq:topological_charge}
\end{equation}%
where $p$ is defined in Eq.~\eqref{eq:p} and the $q$ parameter has been introduced  in Eq.~\eqref{eq:general}. 
We report in Fig.~\ref{fig:solitons} the topological charge densities for $q=1$ and two different winding numbers:
 $n=1$, in the left panel, and $n=2$, in the right panel. In principle, we can obtain solutions with arbitrary large winding numbers, corresponding to a large number of solitons. Notice that having an odd value of $q$ is mandatory to have a non-vanishing topological charge. An odd value  of $q$ implies that the unimodular field $\bm n$ flips at the $\theta$ boundary. One may actually enlarge the $\theta$ boundary, say from $[0,\pi]$ to $[0,2 \pi]$. In this way  the unimodular field $\bm n$ does not flip at the $\theta$ boundary. The resulting solitonic configuration would consist of a crystal of solitons and anti-solitons with vanishing total topological charge. 

\begin{figure}[h!]
\centering
\sidecaption
\includegraphics[width=9.5cm]{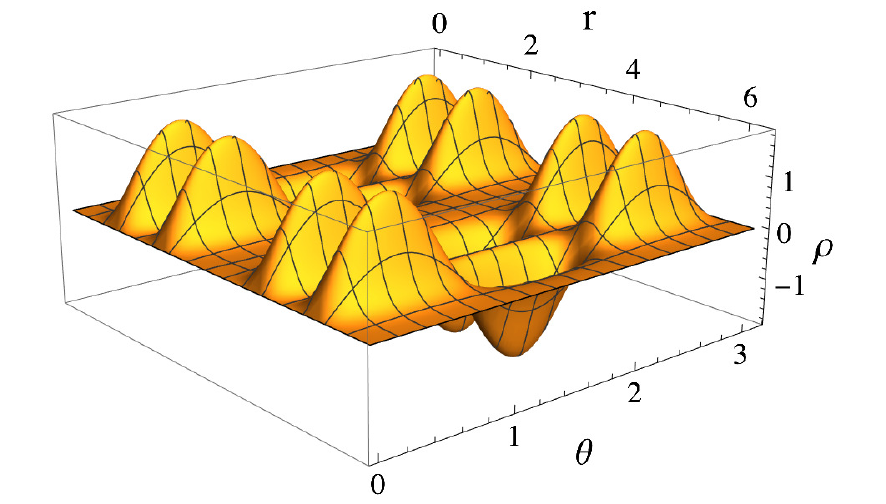}
\caption{Topological density distribution, see Eq.~\eqref{eq:rhom2}, obtained with $q=3$ and  $n=4$. The topological density assumes positive values, corresponding to solitons, as well as negative values, corresponding to anti-solitons.   }
\label{fig:sol_antisol}
\end{figure}

Different solutions arise for values of $q$ different from $1$. For instance, in Fig.~\ref{fig:sol_antisol} we report the solution obtained with $q=3$ and $n=4$. The system now consists of a grid of solitons and anti-solitons.

Since the topological charge  can be understood as  the baryonic charge of the system~\cite%
{Callan:1983nx,Piette:1997ny}, we have obtained a configuration in which  baryons (and eventually anti-baryons) at fixed space points are embedded in a gas of pions. This means that we are actually probing the region of the QCD  phase diagram in Fig.~\ref{fig-1} at non-vanishing $\mu_B$. The  baryonic number density is given by
\be
n_B = \frac{n}{V} \,,
\ee
where $V$ is the volume. For $b=1$, see Eq.~\eqref{eq:ranges} and \eqref{eq:ell}, we have 
\be
V = 4 \pi^3 \ell^3 = \frac{4 \pi^3}{m_\pi^3} \simeq 387 \,\text{fm}^3 \,,
\ee
which is a pretty large volume for baryonic matter. It follows that
\be
n_B \simeq 2.6\, n \times   10^{-3} \, \text{fm}^{-3}\,,
\ee
which for small $n$ is rather low. For instance,   to reach the nuclear saturation density it is necessary  to have $n\simeq 60$ baryons within the volume $V$. 
Finally, we note that a non-vanishing baryonic density implies a non-vanishing baryonic chemical potential.
The fact that the baryonic chemical potential is non-zero is not in contrast with the definition of the  covariant derivative in the $\chi$PT Lagrangian, see Eqs.~\eqref{eq:covariant} and \eqref{eq:deftilda}, because pions have zero baryonic charge, see also the discussion in \cite{Mammarella:2015pxa}. 

% - - - - - - - - - - - - - - - - - - - - - - - - - - - -
\section{Conclusions and outlook}
\label{sec:conclusions} 
% - - - - - - - - - - - - - - - - - - - - - - - - - - - -
We have determined an inhomogeneous solution of a system of pions at non-vanishing isospin chemical potential. 
Remarkably, the isospin chemical potential and the system size are related by Eq.~\eqref{eq:p}, stating that the box length should be proportional to $1/\mu_I$.  
For appropriate boundary conditions, the solutions are characterized by a non-vanishing topological number,  corresponding to the baryonic number of the system. 

Regarding the stability of the obtained solution, the fact that a solitonic configuration has a non-vanishing topological charge does not in general prevent its decay to the standard vacuum.  It is well known that   in three or fewer space-time dimensions the thermal fluctuations can melt  the condensates with one dimensional modulation~\cite%
{Baym:1982ca}. This is indeed  the reason why we have considered a (3+1)-dimensional
modulation. On the other hand, the Derrick's scaling argument~\cite{Derrick:1964ww} implies that  in flat, topologically trivial (3+1)-dimensional space-time, the (3+1)-dimensional modulations are unstable. We have circumvented this theorem considering a finite spatial volume with non-trivial
boundary conditions, which makes the Derrick's scaling argument inapplicable.

Although the solitonic solution  we have found  is not trivially unstable,   it has been determined by the particular ansatz in Eq.~\eqref{eq:general}, therefore it should correspond to a stationary solution which may be   metastable or even unstable to low-energy fluctuations. To analyze its stability we are currently working on   the fluctuations around the background configuration,
\be
\alpha \to  \alpha + \hat \alpha \,, \qquad 
\Theta \to \Theta + \hat \Theta \,, \qquad
\Phi \to \Phi + \hat \Phi \,,
\ee
where $\alpha$, $\Theta$ and $\Phi$ indicate the background solutions and  $\hat \alpha,\hat \Theta$ and $ \hat \Phi $ are the fluctuations. The stability requirement is  that upon inserting the above expressions in Eq.~\eqref{eq:Lgen}  all these fields are well behaved, in particular none of them should have  a  tachyonic mass. The main technical problem in conducting this analysis  is that these fluctuations are non-linearly interacting.  This  analysis is presently underway.

\end{document}